# Effect of local anisotropy on fatigue crack initiation in a coarse grained nickel-base superalloy

*Benedikt Engel*[1*], *Tilmann Beck*[1], *Nadine Moch*[2], *Hanno Gottschalk*[2] and *Sebastian Schmitz*[3]
[1]Institute of Materials Science and Engineering, University of Kaiserslautern, Germany
[2]Bergische Universität Wuppertal, Gaußstraße 20, 42119 Wuppertal, Germany
[3]Siemens AG, Huttenstraße 12, 10553 Berlin, Germany

**Abstract**. In the present work, theoretical approaches, based on grain orientation dependent Young's modulus and Schmid factor are used to describe the influence of local grain orientation on crack initiation behaviour of the coarse grained nickel base superalloy René80. Especially for strongly anisotropic crystal structures with large grain size, such as the investigated material, the local elastic properties must be taken into account for assessment of fatigue crack initiation. With an extension of Schmid's law, the resulting shear stress amplitude, which triggers local cyclic plastic deformation, can be calculated depending on local Young`s modulus and Schmid factor. A Monte Carlo simulation with 100,000 samples shows how random grain orientation affects these parameters. Furthermore, the product of Young`s modulus and Schmid factor (called E·m) is used as a parameter to determine how grain orientation influences resulting shear stress amplitude for given total strain amplitude. In addition to the theoretical work using that approach, this model is also validated using isothermal LCF experiments by determining local grain orientation influence on the crack initiation site using SEM-EBSD analyses.

## 1 Introduction

Due to the increase in the share of regenerative energy in electricity production and the associated decrease in use of fossil sources, i.e. coal and natural gas, it is necessary to compensate the gap between current electricity demand and available renewables [1]. In order to react quickly to these gaps, gas turbines are used, which can be cold started and ramped up to maximum power in short time. While conventional power plants, such as coal or nuclear power plants, require considerable time to generate maximum output from cold start or idle conditions, modern gas turbine power plants, even in the 300 MW class, can achieve this in less than 30 minutes [2]. Hence, frequent starts and stops as well as load changes result in thermally induced low-cycle-fatigue (LCF) loadings of the turbine blades made of nickelbase superalloys, which can become more critical for component life than creep loads. Therefore, a fundamental understanding of the crack initiation and fatigue process of these materials becomes increasingly important. Current lifetime prediction models use classic deterministic approaches considering high safety factors and hence, less efficiency is achieved in turbine design than possible in case of optimal usage of the Ni-alloys' fatigue strength. Recent design approaches therefore use probabilistic models for lifetime calculation with improved material models. For this purpose, a significantly higher number of influencing parameters must be taken into account to describe fatigue crack initiation behaviour. In this context, the local grain orientation at the crack initiation site is of particular interest. Due to the large grain size of conventionally cast Ni-base turbine blades, the regions of the component undergoing highest cyclic stress only contain few grains. The random orientation of these grains leads to locally different elastic and plastic material properties significantly influencing the crack initiation behaviour. Related to Schmid`s law for single crystals or single grains, the resulting shear stress in the slip systems depends on the applied normal stress as well as the orientation dependent Schmid factor. If the local shear stress exceeds the critical resulting shear stress (CRSS), dislocation slip becomes effective and, in case of cyclic loading, a fatigue crack may be initiated. Moreover, at a given integral normal stress, e.g. in a uniaxially loaded test piece, the local normal stress to be taken into account in Schmid`s law also depends on local elastic grain stiffness, which again is a function of grain orientation. Nickel-based materials have a high anisotropy factor of more than 2.5. Hence, Young`s modulus can vary from 130 GPa to 330 GPa [3,4] depending on grain orientation, which has a significant influence on the resulting shear stresses in the slip systems.

Strain-controlled isothermal LCF tests at conventionally cast Ni-Superalloys typically show large fatigue lifetime scatter up to a factor of 10, even in the LCF regime [5–9]. A major reason for this are orientation dependent resolved shear stresses in the randomly oriented coarse grains. Previous work has shown the influence of the Schmid factor on lifetime behaviour by calculating a resulting shear stress – lifetime diagram by multiplying the integral stress amplitude measured at half lifetime in strain controlled LCF-tests with the Schmid factor of the crack initiation grain. The resulting CRSS vs. Number of cycles to crack initiation diagram showed clearly reduced scatter compared with the representation total strain amplitude vs. number of cycles to failure [5,6,10]. However, these results and EBSD measurements also show that fatigue cracks are not necessarily initiated in slip systems with the highest Schmid factor at the specimen surface [10].

---

[*] Corresponding author: engel@mv.uni-kl.de

Therefore, crack initiation prediction only based on the Schmid factor is obviously not covering all major influences on fatigue of coarse grained polycrystalline Ni-base superalloys. Research on HCF of Ni-base superalloys with grain sizes in the µm range has already shown that both, Young's modulus and Schmid factor including their interrelation via grain orientation must be considered to predict the crack initiating grains. In this context, Pollock et al. showed that crack initiation in René88DT, manufactured by powder metallurgy, is dependent on the occurrence of twins in addition to Schmid factor and Young's modulus. For this material cracks often initiated in grains with slip systems oriented parallel to twin boundaries [11–13].

## 2 Investigated Material

The investigated material is the polycrystalline Ni-base superalloy René80 frequently used for gas turbine blades [13–15]. Typical operation temperatures are between 760 °C and 982 °C. The material contains approximately 30 % - 35 vol.-% of cuboidal primary γ'- particles with an average size of 0.4 µm, measured by scanning electron microscopy, semi-coherently embedded in the fcc γ – matrix. In addition, spherical secondary γ' is observed within the nickel matrix. The specimen were delivered as 150 mm long casted bars, with diameters of 20 mm. The chemical composition measured by the manufacturer is given in Table1.

**Table 1.** Chemical composition of René80 (in weight %)

| Element | Composition |
|---|---|
| Ni | balance |
| Cr | 14.04 |
| Co | 9,48 |
| Ti | 5.08 |
| Mo | 4.03 |
| W | 4.02 |
| Al | 2.93 |
| C | 0.17 |
| B | 0.0151 |
| Zr | 0.011 |

LCF-tests at 850 °C were carried out using a 100 kN servo-hydraulic testing system (MTS type 810) with inductive heating (Hüttinger TIG 5/300). For temperature control a Eurotherm 2604 device was used combined with a ribbon-type thermocouple (type K) attached in the middle of the gauge-length. All isothermal LCF tests were performed in total strain control with fully reversed loading. A 12mm MTS high temperature extensometer was used for strain measurement. The temperature gradient within the cylindrical centre section of the sample (length 18mm, diameter 7mm), measured with three ribbon-type thermocouples, does not exceed ± 8°C within a distance of +/- 5 mm from the middle of the gauge length.

## 3 Description of anisotropic material behaviour

### 3.1 Elastic Anisotropy

Due to the crystalline structure of metallic materials, elastic properties are in most cases direction dependent within a single grain. A non-textured polycrystal is composed of statistically randomly oriented grains, separated from each other by grain boundaries. If a sufficiently high number of randomly oriented grains exists in the gauge length of a specimen or within technical component, the anisotropic elastic properties of the individual grains are averaged over the high grain number and so-called quasi-isotropic material behaviour can be assumed in good approximation. Hooke's law can then be used in its simple form representing a linear correlation between stress and strain for isotropic material behaviour. This assumption becomes questionable if only a small number of grains in a specimen gauge length or component are to be taken into account, especially in case of materials with high elastic anisotropy factor. An isotropic approach would then result in a large scatter of mechanical properties including fatigue strength and, hence, local grain properties need to be taken into account. The following section shows how the elastic properties of a crystallite can be determined for any given orientation.

If a general linear elastic material law is considered, the two second order stress and strain tensors are combined with a fourth rank stiffness tensor as shown in equation 3.1, known as generalized Hooke's law [14].

$$\varepsilon_{ij} = S_{ijkl}\sigma_{kl} \quad (3.1)$$

For a completely anisotropic material, the stiffness matrix has 81 independent components. For a cubic lattice, three independent stiffness components $S_{11}, S_{12}, S_{44}$ exist due to rotational planes in the cubic system and symmetry of stress and strain tensor. Hence, using equation 3.2, an anisotropic cubic material can be described completely.

$$S_{ij} = \begin{pmatrix} S_{11} & S_{12} & S_{12} & & & \\ S_{12} & S_{11} & S_{12} & & & \\ S_{12} & S_{12} & S_{11} & & & \\ & & & S_{44} & & \\ & & & & S_{44} & \\ & & & & & S_{44} \end{pmatrix} \quad (3.2)$$

To determine the stiffness as a function of grain orientation, the stiffness tensor must be rotated four times with a rotational matrix U (eq. 3.4) [15–17].

$$S'_{ijkl} = U_{ig}U_{jh}S_{ghmn}U_{km}U_{ln} \quad (3.3)$$

U represents a rotational matrix containing the rotation angles around the default crystal coordinate axes. A common convention for this is the Bungee notation with the Euler angles $\varphi_1$, $\varphi_2$ and $\theta$. A detailed description of the Bungee notation and the Euler angles can be found in [17]. Each rotation can be described with its own rotational matrix and multiplication of all three rotational matrices results in the complete rotational matrix U (3.4), with s,c for sine and cosine.

$$U = \begin{bmatrix} c(\varphi_1)c(\varphi_2) - s(\varphi_1)c(\theta)s(\varphi_2) & -c(\varphi_1)s(\varphi_2) - s(\varphi_1)c(\theta)c(\varphi_2) & s(\theta)s(\varphi_1) \\ s(\varphi_1)c(\theta) + c(\varphi_1)c(\theta)s(\varphi_2) & -s(\varphi_1)s(\theta) + c(\varphi_1)c(\varphi_2) & -s(\theta)c(\varphi_1) \\ s(\theta)s(\varphi_2) & s(\theta)c(\varphi_2) & c(\theta) \end{bmatrix} \quad (3.4)$$

Equations commonly used in literature allow an analytical calculation of effective Young`s Modulus in normal stress direction [4,17–19]. The equation used here consists of the three independent stiffness components $S_{11}$, $S_{12}$ and $S_{44}$ combined with an orientation factor r.

$$\frac{1}{E_{hkl}} = S_{11} - 2\left(S_{11} - S_{12} - \frac{1}{2}S_{44}\right)r_{\{hkl\}} \quad (3.5)$$

The orientation factor r can be expressed with the Euler angles $\varphi_1$, $\varphi_2$ and $\theta$ by the following equations for stresses in z-direction.

$$r_{\{hkl\}} = g_{13}^2 g_{23}^2 + g_{23}^2 g_{33}^2 + g_{33}^2 g_{13}^2 \quad (3.6)$$

with

$$g_{13} = \sin\theta \sin\varphi_2, \; g_{23} = \cos\varphi_2 \sin\theta, \\ g_{33} = \cos\theta \quad (3.7)$$

Figure 1 shows the geometric dependency of orientation and r factor in a pole figure for cubic materials.

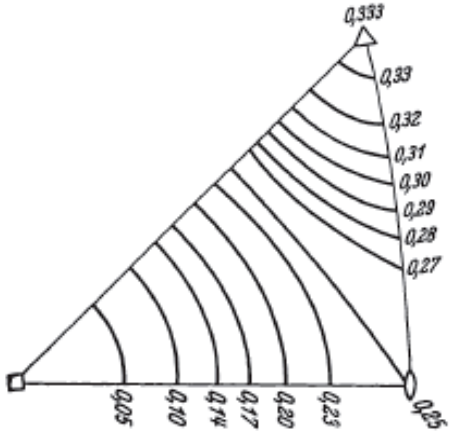

**Fig. 1.** Dependence of grain orientation on the geometrical r-factor [20].

Equation 3.6 corresponds with the bottom line of the rotational matrix U and shows the independence of $\varphi_1$ on the calculation of the Young's modulus due to its invariant rotation around the stress axis z.

### 3.2 Plastic Anisotropy

As a prerequisite of plastic deformation, it is assumed that the shear stress in the slip systems resulting from the applied stress must exceed a critical shear stress, which can be considered as a material-specific threshold value. To calculate the resulting shear stress a more generalized, Schmid-law is used [21–24].

$$\tau_{res} = s_{ij} \cdot \sigma \cdot n_i \quad (3.8)$$

Equation 3.8 shows the calculation of resulting shear stress where $\sigma$ represents a given stress tensor, $s_{ij}$ the normalised slip plane direction vector and $n_i$ the normalised slip plane normal vector. With this equation, the resulting shear stress can be calculated based on the stress tensor in each slip system (or in every other crystallographic system of interest). For a rotation of the grain or elementary cell, the slip systems also rotates. For this purpose, the plane and directional vectors are rotated with the rotation matrix U mentioned in equation 3.4. Furthermore, the value of $\sigma_{33}$ is set to 1 for uniaxial loading (in z direction) to calculate the Schmid factor. Plastic deformation can be expected to occur first in slip systems with the highest resulting shear stress, i.e. the highest Schmid factor ($\max(\tau_{res})$).

$$\tau_{res} = U \cdot n_i \cdot \sigma \cdot U \cdot s_{ij} \quad (3.9)$$

$$\text{and } \sigma = \begin{bmatrix} 0 & 0 & 0 \\ 0 & 0 & 0 \\ 0 & 0 & 1 \end{bmatrix} \quad (3.10)$$

### 3.3 Combination of Schmid factor and Young's modulus

To illustrate the correlation between Schmid factor and Young`s modulus, a single grain in stress controlled loading is considered. The material is assumed to be isotropic and uniaxial load is applied. In this case the resulting shear stress amplitude can be determined using Schmid's law.

$$\tau_{a,res} = \sigma_a \cdot m \quad (3.11)$$

In the next step, the load condition is changed to a total strain control. Assuming approximately elastic material behaviour, the stress amplitude σ can be estimated conservatively from Hooke's law and be inserted into Eq. 3.11:

$$\tau_{a,res} = \varepsilon_{a,t} \cdot E \cdot m \quad (3.12)$$

Equation 3.12 clarifies that, in case of approximately elastic behaviour, the resulting shear stress in each considered grain only depends on the product of Young's modulus and Schmid factor. And as shown above, both, Schmid factor and Young's modulus depend on the same grain orientation while $\varepsilon_{a,t}$ is constant for strain controlled tests. Note that equation 3.12 is exactly valid only until the resulting shear stress exceeds the critical shear stress within the slip system. From this point on, plastic deformation occurs which is no longer associated with the linear assumption of Hooke's law and the resulting shear stress is overestimated by eq. 3.12.

## 4 Generation of random orientation

To investigate the correlation of Young`s modulus and Schmid factor depending on grain orientation, a Monte Carlo Simulation with 100.000 samples was performed. To generate random orientations a normal vector placed on the [001] plane of a face-centered cubic cell is taken, located in the middle of a three-dimensional coordinate system. By rotating the cubic cell, the vector is oriented to different points on the surrounding spherical surface. With sufficiently high amount of random orientations, the whole surface is sufficiently covered and each point on this sphere representing an orientation of the cubic cell. Due invariant rotation around z-axis, because of uniaxial loading, the sphere can be parameterized with two Euler angles. The assumption of a uniform distribution of these angles leads to an accumulation of points at the corresponding poles, as the sphere and its projection in figure 2 a) shows.

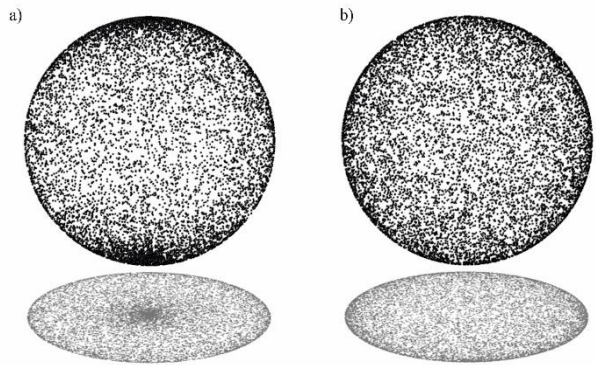

**Fig. 2.** 10.000 randomly distributed points on a sphere. a) Uniform distribution of the parameters b) homogeneous distribution achieved by Haar measure.

The distribution in a), with increased orientations in [001]- directions, would lead to misleading results in frequency distributions of schmid factor and Youngs modulus. To achieve a homogeneous distribution, the three dimensional Haar measure is projected onto the two dimensional sphere. Based on this distribution, which is shown in figure 2b), 100.000 random SO3 matrices were generated to simulate a statistically random orientation of the grain.

## 5 Results and Discussion

### 5.1 Monte Carlo Simulation

Due to lack of available materials data for René80, the calculations were performed instead based on data of INC738 LC, which is very similar in chemical composition and γ/γ' structure. The following Monte Carlo simulation for the distribution of Young`s modulus was made for a single INC738 LC grain at 25°C and 850°C, respectively. The elastic constants for a wide temperature range were taken from [26]. For 25 °C the values are $S_{11} = 0.00797$, $S_{12} = -0.00323$ and $S_{44} = 0.00787$. The stiffness constants used in equation 1.5 and the Euler angles for the rotation factor were extracted from the rotational matrices according to equation 1.4. Figure 3 summarizes the results as a histogram.

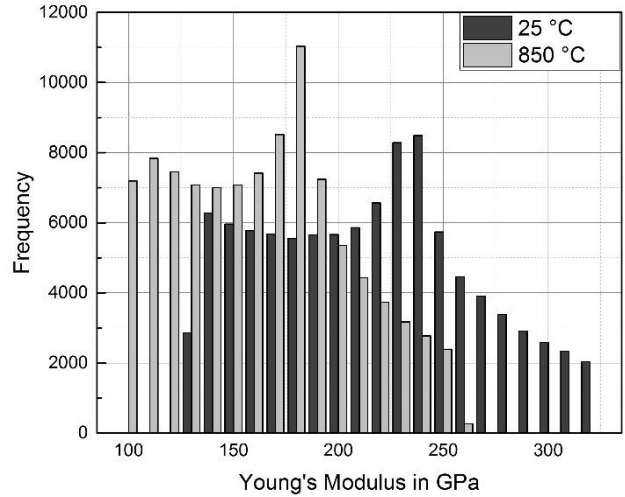

**Fig. 3.** Result of the Monte Carlo Simulation for 100.000 random orientations. INC738 at 25 °C and 850 °C.

Due to elastic anisotropy, Young`s modulus ranges from 130 GPa up to 330 GPa at 25°C with a maximum density at 240 GPa representing the high anisotropy factor. The mean value is 213 GPa. A second simulation for 850 °C shows a similar distribution, but shifted to lower values due the temperature influence on Young`s modulus. The elastic constants for INC738 LC at 850 °C are determined by linear interpolation of the values from [26] between 800 °C and 900 °C, resulting in $S_{11} = 0.010963$, $S_{12} = 0.00457$ -and $S_{44} = 0.010122$. Notably, the range of Young`s Modulus is reduced compared 25°C leading to higher frequencies in the distribution at low Young's moduli. The maximum density of Young`s moduli for 850 °C is found to occur at values 180 GPa, with a more pronounced peak in frequency compared to room temperature. The mean value is 158 GPa. The occurrence of increasing frequencies with increasing temperature can be explained by figure 4. For orientations close to [111] Young's Modulus decreases by 80 GPa from 25 °C to 900°C. For the same temperatures the Young's moduli near [100] orientation change only by 38 GPa. As a result, the distribution of Young's Moduli is "compressed" at higher temperatures, leading to a higher number density in the frequency distribution.

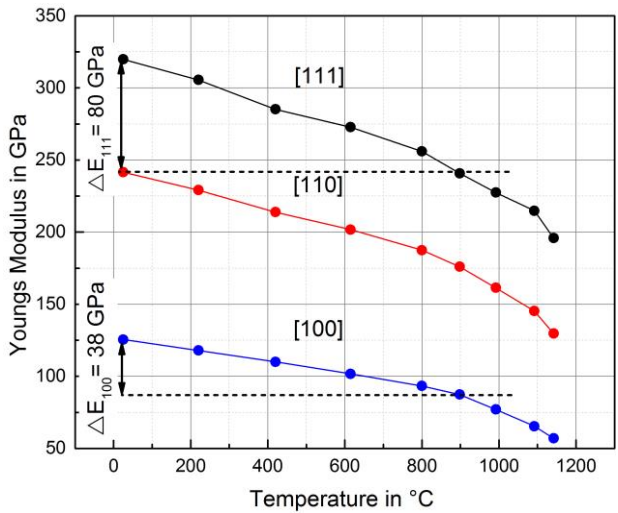

**Fig. 4.** Temperature Dependence of Young's modulus in selected lattice directions of INC738 LC calculated from data according to [7].

Opposed to Young's modulus, the Schmid factors for the considered slip systems are temperature independent because they result entirely from geometric considerations, taking only the crystallographic structure of the material and the vectors of the slip systems into account. The face centred crystallographic structure of nickel base superalloy leads to 12 primary slip systems of the type {111}[110]. To determine the highest Schmid factor, eq. 2.2 is used to calculate the Schmid factors for all slip systems in dependence of the rotational matrices, while only the highest is taken into account in the histogram in Fig. 5.

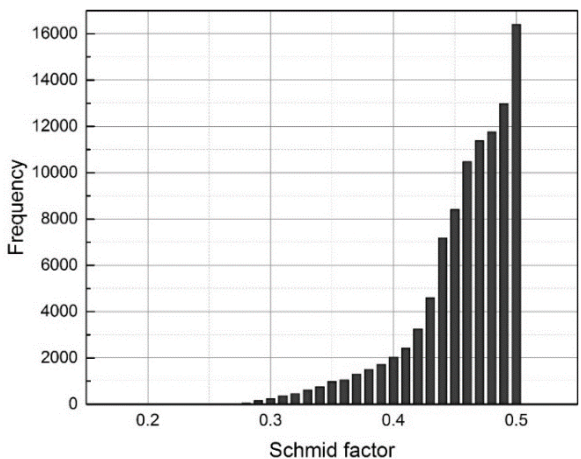

**Fig. 5.** Schmid factor distribution for 100.000 random grain orientations of a face centred cubic cell.

It can be seen that the frequencies in the histogram increase up to the maximum Schmid factor of 0.5. The reason for this is the large number of slip systems, due to which, for random orientation, a high probability exists for at least one of the twelve slip systems to feature a high Schmid factor. Only a few orientations show low maximum Schmid factors. These orientations correspond to those where the load axis direction approximately corresponds to the [111] directions of the elementary cell. The Average Schmid factor of 100.000 random orientations is 0.452 with a standard deviation of 0.0413.

## 5.1 E·m - Modell

The resulting shear stress in each slip system of a uniaxial loaded, anisotropic grain only depends on the product of Young's modulus and Schmid factor, in the following referred to as E·m. Figure 6 shows E plotted against the Schmid factor as a result of a Monte Carlo simulation for 25°C in one diagram. The product E·m is indicated by a greyscale coding, while each point corresponds to a single, randomly chosen orientation.

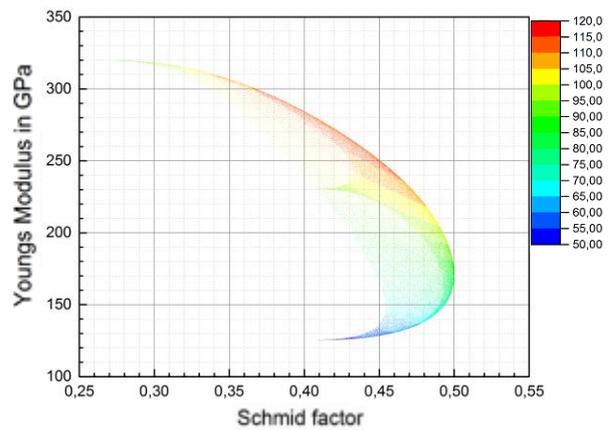

**Fig. 6.** Distribution of Schmid factor and Young's modulus calculated for IN738 LC at T = 25°C with 100.000 random grain orientation

The distribution (Fig.6) shows highest shear stresses, i.e. highest E·m values for orientations with Schmid factors between 0.37-0.45 and high Young's moduli, from 240 GPa up to 300 GPa. As a result of the frequency distributions discussed above, these combinations occur less frequently while orientations with high Schmid factors up to 0.5 appear more often. It should be noted that these orientations correlate with low Young's moduli of 130 GPa - 200 GPa. A second simulation at 850 °C show that in spite of a different distribution of Young's moduli (see Fig. 3) the E·m distribution remains similar as for room temperature. Highest shear stresses occur for orientations with Schmid factors from 0.37 – 0.45 and corresponding Young's moduli of 190 GPa – 240 GPa.

Further visualisation of the behaviour of Young's modulus and Schmid factor depending on grain orientation is possible as a three-dimensional body (Fig. 7). In the centre of a considered sphere is a cubic cell. As described above, for random orientations of the cell, the [001] vector is directed to the sphere surface in dependence of its rotation. To demonstrate the influence of E·m in dependence of grain orientation, the [001] vector is rotated with the relevant rotational matrix and multiplied with the value of E·m. The distance from the centre to a point on the three-dimensional body`s surface represents the E·m value for the randomly chosen orientations described in section 4, in case of uniaxial loading in [001]-direction. For better illustration, only one half of the axisymmetric body is plotted in Fig. 7.

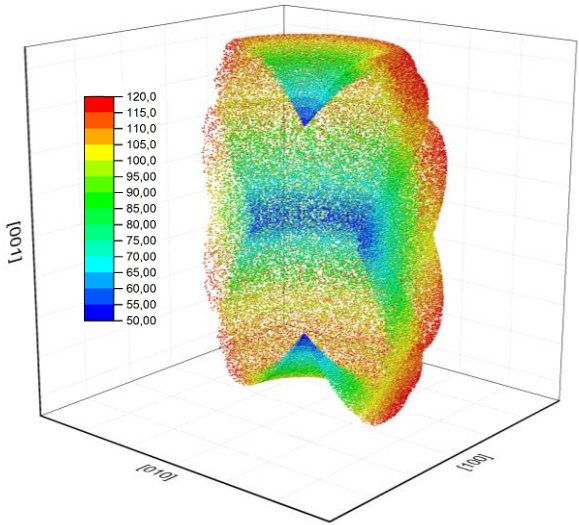

**Fig. 7.** Three dimensional representation of orientation dependence of E·m for uniaxial stress in [001] direction. Calculated for 200.000 orientations of INC738 LC at 25°C.

Besides visualizing the grain orientation dependency of E·m, Fig. 7 underlines the excellent properties of [001] directed single crystalline (SX) or directionally solidified (DS) Ni-base alloys: Orientations near [001] lead to lowest Young's moduli combined with moderate Schmid factors which results in a global minimum of E·m. Moreover, it becomes apparent that even small manufacturing induced changes in crystal orientation cause a steep increase in E·m, which detrimentally effect e.g. thermo-mechanical fatigue behaviour of turbine blades manufactured from SX and DS Ni-Base alloys, respectively.

## 5.2 Experimental Validation of E·m Modell

To validate the theoretically determined correlation between E·m and fatigue crack initiation, cracked specimen from isothermal LCF experiments were investigated. Because of the relatively low total strain amplitudes, single crack initiation within individual grains was observed [9] and, according to the nearly closed stress strain hysteresis loops recorded during the tests, globally elastic behaviour can be assumed in very good approximation. Due formation of a single crack, it can be assumed that only one grain has exceeded the critical shear stress during the test and plastic deformation occurred on local scale. This leads to the formation of persistent slip bands in the relevant grain, inducing extrusions and intrusions at the adjacent surface. The resulting micro notch effect leads to an increase in local stress and finally to crack initiation [10,27,28]. To investigate the grain orientation in the vicinity of the crack origin via EBSD, the specimen were cut longitudinaly, grinded, mechanically and electrolytically polished. Figure 8 shows a longitudinal cut through the region around the crack initiation grain (grain 1) of two exemplarily chosen specimens.

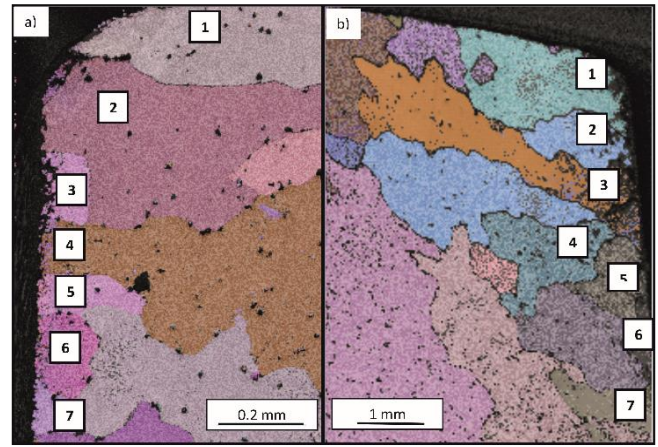

**Fig. 8.** EBSD Measurement for the crack initiation grain (1) and surrounding grains.

**Table 2.** Calculated Young's modul (for T = 850 °C), Schmid factors and E·m values.

| Grain Nr. a) | Young's Moduli | Schmid factor | E·m |
|---|---|---|---|
| 1 | 152 | 0.48 | 73.2 |
| 2 | 95 | 0.46 | 44.2 |
| 3 | 97 | 0.46 | 45.8 |
| 4 | 95 | 0.45 | 42.8 |
| 5 | 97 | 0.46 | 45.2 |
| 6 | 102 | 0.48 | 49.1 |
| 7 | 102 | 0.47 | 48.7 |
| Grain Nr. b) | Young's Moduli | Schmid factor | E·m |
| 1 | 167 | 0.45 | 76.48 |
| 2 | 173 | 0.46 | 81.61 |
| 3 | 178 | 0.43 | 77.60 |
| 4 | 106 | 0.48 | 52.05 |
| 5 | 96 | 0.46 | 44.95 |
| 6 | 128 | 0.49 | 63.74 |
| 7 | 143 | 0.49 | 70.47 |

The calculation of local Young's moduli and Schmid factors (Table 2) for the same specimens was carried out using the measured Euler angles $\varphi_1$, $\varphi_2$ and $\theta$. In specimen a), both, grain 1 and 6 feature the highest Schmid factors of m = 0.48. If crack initiation would be assumed to occur in grains with high Schmid factors, the crack could also have started in grain 6. However, grain 1 provides a significantly higher Young's modulus which leads to higher E·m values and therefore to local plastic deformation and crack initiation in grain 1. Considering sample b), it becomes apparent that fatigue failure did not start in the grain with highest E·m, but, the crack initiation grain 1) and the surrounding grains show higher values of E·m than the other grains. Thus crack initiation starts in this case at a grain agglomerate with high E·m values. Other feature significantly higher Schmid factors (Grain 6, 7) but lower shear stresses are induced due to the lower Young's moduli.

## 6 Concluding remarks

It could be shown that especially for elastic anisotropic materials such as nickel-base superalloys, that, besides the influence of the Schmid factor investigated in earlier studies [10, 21, 23], local elastic stiffness has an important impact on crack initiation. If homogenous strain is assumed in the considered region, the shear stress in the slip system depends linearly on the product of Schmid factor and Young's modulus. As soon as this resulting shear stress exceeds a critical value, plastic deformation, or, in case of cyclic loading, fatigue crack initiation may be initiated. A Monte Carlo simulation with 100.000 samples showed the frequency distributions of Young's modulus as well as Schmid factor and revealed orientations with Schmid factors in the range 0.37-0.45 and Young's moduli of about 240 GPa up to 300 GPa to be most critical at room temperature. These orientations induce highest shear stresses at a given uniaxial strain but occur less frequently. Orientations with maximum Schmid factor occur much more frequently but induce significant lower shear stresses within the slip systems because of lower Young's moduli. A similar behaviour was shown for T = 850 °C. The influence of both, Young's modulus and Schmid factor on the crack initiation site in coarse grained Ni-base superalloys was experimentally proven for isothermal total strain controlled fatigue tests with single crack initiation. EBSD measurements of cracked LCF specimens showed crack initiation in grains with orientations featuring intermediate Schmid factors and high Young's moduli and therefore high E·m or crack initiation in grain accumulation with high E·m. With knowledge of the elastic constants and the investigated crystal system, the E·m model can be applied to several materials. However, it should be noted that the presented failure model in its actual state only applies to uniaxial loadings and globally homogenous strain distribution is assumed. Ongoing research deals with the extension of the E·m model to multiaxial loading conditions as well as with FEM-based consideration of stress/strain inhomogeneities caused by interaction between differently oriented grains.

The investigations were conducted as part of the joint research programme COOREFLEX TURBO the frame of AG Turbo. The work was supported by the Bundesministerium für Wirtschaft und Energie (BMWi) as per resolution of the German Federal Parliament under grant number 03ET7041K. The authors gratefully acknowledge AG Turbo and Siemens AG for their support and permission to publish this paper. The responsibility for the content lies solely with its authors.